\documentclass[preprint,aps]{revtex4}
\usepackage{amsfonts}
\usepackage{amsmath}
\usepackage{amssymb}
\usepackage{graphicx}

\newcommand{\be}{\begin{equation}}
\newcommand{\ee}{\end{equation}}
\newcommand{\ben}{\begin{eqnarray}}
\newcommand{\een}{\end{eqnarray}}
\newcommand{\ra}{\rangle}
\newcommand{\la}{\langle}
\newcommand{\m}{\hbox{-}}

\newcommand{\rra}{\right\rangle}

\newcommand{\lli}{\left|}
\newcommand{\rli}{\right|}

\begin{document}

\title{Entropic Entanglement Criteria for Fermion Systems}
\author{C. Zander$^1$, A.R. Plastino$^{2,3}$\footnote{E-mail:arplastino@ugr.es},
M. Casas$^4$ and A. Plastino$^3$} \affiliation{$^1$Physics Department, University of Pretoria -
Pretoria 0002, South Africa\\ $^2$Instituto Carlos I de F\'isica Te\'orica y
Computacional, University of Granada, Granada, Spain, EU\\
$^3$National University La Plata, UNLP-CREG-IFILP-CONICET - C.C. 727, 1900 La Plata, Argentina\\
$^4$Departament de F\'isica, Universitat de les Illes Balears and
IFISC-CSIC, 07122 Palma de Mallorca, Spain }

\begin{abstract}
Entanglement criteria for general (pure or mixed) states of
systems consisting of two identical fermions are introduced. These
criteria are based on appropriate inequalities involving the
entropy of the global density matrix describing the total system,
on the one hand, and the entropy of the one particle reduced
density matrix, on the other one. A majorization-related relation
between these two density matrices is obtained, leading to a
family of entanglement criteria based on R\'enyi's entropic
measure. These criteria are applied to various illustrative 
examples of parametrized families of mixed states.
The dependence of the entanglement detection efficiency
on R\'enyi's entropic parameter is investigated. The extension of
these criteria to systems of $N$ identical fermions is also
considered.

\vskip 0.5cm


\end{abstract}
\maketitle

\section{Introduction}

The entanglement features exhibited by systems consisting of
identical fermions have attracted the attention of several
researchers in recent years
\cite{SCLL01,ESBL02,AFOV08,GM04,GMW02,LNP05,NV07,BBB07,BPCP08,OSTS08,LV08,ZP10,ZZX02,YPD09,PMD09}.
Entanglement in fermion systems has been studied in connection
with different problems, such as the entanglement between
electrons in a conducting band \cite{NV07}, the entanglement
dynamics associated with scattering processes involving two
electrons \cite{BBB07}, the role played by entanglement in the
time-optimal evolution of fermionic systems \cite{BPCP08,OSTS08}, the
classification of three fermion states based on their entanglement
features \cite{LV08}, the detection of entanglement in fermion
systems through the violation of appropriate uncertainty relations
\cite{ZP10}, the entanglement features of fractional quantum 
Hall liquids \cite{ZZX02} and the entanglement properties of the 
eigenstates of soluble two-electrons atomic models \cite{YPD09}.

The concept of entanglement in systems of indistinguishable
particles exhibits some differences from the corresponding concept
as applied to systems consisting of distinguishable parts. There
is general consensus among researchers that in systems of
identical fermions the minimum quantum correlations between the
particles that are required by the antisymmetric character of the
fermionic state do not contribute to the state's amount of
entanglement
\cite{SCLL01,NV07,GM04,GMW02,LNP05,ESBL02,AFOV08,ZZX02,YPD09,BBB07,BPCP08,OSTS08,LV08,ZP10,PMD09}.
This means that the separable (that is, non-entangled) pure states
of $N$ fermions are those having Slater rank $1$. These are the
states whose wave function can be expressed (with respect to an
appropriate single particle basis) as a single Slater determinant
\cite{AFOV08}. On the other hand, the set of mixed non-entangled
states comprises those states that can be written as a statistical
mixture of pure states of Slater rank $1$. Here, when discussing
systems of identical fermions, we are considering entanglement
between particles and not entanglement between modes.

In the case of pure states of two identical fermions, necessary
and sufficient separability criteria can be formulated in terms 
of the entropy of the single particle reduced density matrix
\cite{GM04,LNP05,PMD09}. Alas,
no such criteria are known for general, mixed states of two
fermions, except for the case of two fermions with a single
particle Hilbert space of dimension $4$, for which a closed
analytical expression for the concurrence (akin to the celebrated
Wootters' formula for two-qubits \cite{W98}) is known.  In
general, to determine whether a given density matrix of a
two-fermion system represents a separable state or not is a
notoriously difficult (and largely unexplored) problem.
Consequently, there is a clear need for practical separability
criteria, or entanglement indicators, which can be extended to
systems of higher dimensionality or to scenarios involving more
than two fermions \cite{PMD09}. \\

Entropic separability criteria have played a distinguished role in
the study of the entanglement-related features of mixed states of
multipartite systems constituted by distinguishable subsystems
\cite{HHH96,HH96,HHHH09,BCPP02,VW02,BCPP03,BCPP05}. For this
kind of composite quantum systems, non-entangled states behave
classically in the sense that the entropy of a subsystem is always
less or equal than the entropy of the whole system. If the entropy
of a subsystem happens to be larger than the entropy of the whole
system, then we know for sure that the state is entangled (that is,
this constituteas a sufficient entanglement criteria). This
statement can be formulated mathematically in terms of the R\'enyi
entropic measures,

\be \label{ReyNy}
 S^{(R)}_q [\rho] = \frac{1}{1-q}\ln({\rm Tr}[\rho^q]),
\ee

\noindent
leading to the following family of inequalities satisfied by
separable states \cite{HHH96,HH96,HHHH09,BCPP02,VW02,BCPP03,BCPP05},

\ben \label{distingrenicrit} S^{(R)}_q[\rho_{A}]  &\le &
S^{(R)}_q[\rho_{AB}] \cr S^{(R)}_q[\rho_{B}]  &\le &
S^{(R)}_q[\rho_{AB}]. \een

\noindent In the above equations $\rho_{AB}$ is the joint density
matrix describing a bipartite system consisting of the subsystems
$A$ and $B$, and $\rho_{A,B}$ are the marginal density matrices
describing the subsystems. The entropic parameter in
(\ref{ReyNy}-\ref{distingrenicrit}) adopts values $q \ge 1$. In
the limit $q \rightarrow 1$ the R\'enyi entropy reduces to the von
Neumann entropy. Note that the entropic criteria considered in 
\cite{HHH96,HH96,HHHH09,BCPP02,VW02,BCPP03,BCPP05} and in the
present work, which depend on the entropies of the total 
and reduced density matrices, are different from those 
studied in \cite{G04}, which involve entropic uncertainty
relations associated with the measurement of particular 
observables.

The study of entropic entanglement criteria based upon the above
considerations has been the focus of a considerable amount of
research over the years
\cite{HHH96,HH96,HHHH09,BCPP02,VW02,BCPP03,BCPP05}. It would
be interesting to extend this approach to systems consisting of
identical fermions. The aim of this paper is to investigate
entanglement criteria for general (mixed) states of systems of two
identical fermions based upon the comparison of the entropy of the
global density matrix describing the total system and the entropy
of the one particle reduced density matrix. 
\\

The organization of the paper is as follows. A brief review of
entanglement between particles in systems of identical fermions is
given in Section II. Entropic entanglement criteria for systems of
two identical fermions based on the von Neumann, the linear, and
the R\'enyi entropies are derived in Section III. These entropic
criteria are applied to particular families of states of
two-fermion systems in Sections IV and V. The extension to systems
of $N$ fermions of the entanglement criteria based upon the
R\'enyi entropies is considered in Section VI. Finally, some
conclusions are drawn in Section VII.

\section{Entanglement Between Particles in Fermionic Systems}\label{2}

The concept of entanglement between the particles in system of
identical fermions is associated with the quantum correlations
exhibited by quantum states on top of the minimal correlations due
to the indistinguishability of the particles and the
anti-symmetric character of fermionic states. A pure state of
Slater rank one of $N$ identical fermions (that is, a state that
can be described by one single Slater determinant) must be
regarded as separable (non-entangled) \cite{ESBL02,AFOV08}. 
The correlations exhibited
by such states do not provide a resource for implementing
non-classical information transmission or information processing
tasks. Moreover, the non-entangled character of states of Slater
rank one is consistent with the possibility of assigning complete
sets of properties to the parts of the composite system
\cite{GM04}. Consequently, a pure state of two identical fermions
of the form

\be\label{slaterdet}|\psi_{sl}\ra =
\frac{1}{\sqrt{2}}\lbrace|\phi_1\ra|\phi_2\ra -
|\phi_2\ra|\phi_1\ra \rbrace, \ee

\noindent where $|\phi_1\ra$ and $|\phi_2\ra$ are orthonormal
single-particle states, is regarded as separable.

A pure state $|\psi\ra$ of a system of $N$ identical fermions has
Slater rank $1$, and is therefore separable, if and only if

\be \rm{Tr}(\rho_1^2) = \frac{1}{N}, \ee

\noindent where $\rho_1 =\rm{Tr}_{2, \ldots, N}(\rho)$ is the single
particle reduced density matrix,  $\rho=|\psi\ra\la\psi|$,  $n$ is
the dimension of the single particle state space and $N\le n$
\cite{PMD09}. On the other hand, entangled pure states satisfy

\be \frac{1}{n} \leq \rm{Tr}(\rho_1^2) < \frac{1}{N}.
 \ee

Non-entangled mixed states of systems of $N$ identical fermions are
those that can be written as a mixture of Slater determinants,

\be\label{mixedsep} \rho_{sl} = \sum_i \lambda_i
|\psi_{sl}^{(i)}\ra\la\psi_{sl}^{(i)}|, \ee

\noindent where the states $|\psi_{sl}^{(i)}\ra $ can be expressed
as single Slater determinants, and $0\le \lambda_i \le 1$ with
$\sum_i \lambda_i = 1$.

 Systems of identical fermions with a single-particle
 Hilbert space of dimension $2 k$ (with $k \ge 2$)
can be formally regarded as systems consisting of spin-$s$
particles, with $s =(2k-1)/2$. The members $\{|i\rangle, \,\,
i=1,\ldots, 2k\}$ of an orthonormal basis of the single particle
Hilbert space can be identified with the states $|s, m_s \rangle
$, with $m_s=s-i+1, \,\, i=1, \ldots, 2k$.  We can use for these
states the shorthand notation $\{|m_s\rangle, \,\, m_s=-s, \ldots,
s\}$, because each particular example discussed here will
correspond to a given value of $k$ (and $s$). According to this
angular momentum representation, the antisymmetric joint
eigenstates $\{ |j, m\rangle, \,\, -j\le m \le j, \,\, 0\le j\le
2s \}$ of the total angular momentum operators $J^2$ and $J_z$
constitute a basis for the Hilbert space associated with a system
of two identical fermions. The antisymmetric states $|j, m\rangle$
are those with an even value of the quantum number $j$.
\\

A closed analytical expression for the concurrence of general (pure
or mixed) states of two identical fermions sharing a single particle
Hilbert space of dimension $4$ (corresponding to $s=3/2$) was
discovered by Eckert, Schliemann, Bruss, and Lewenstein (ESBL) in
\cite{ESBL02}. The ESBL concurrence formula is

\be {\mathcal C}_{{\mathcal F}}(\rho) = {\rm max}\{0,\lambda_1 -
\lambda_2- \lambda_3 - \lambda_4 - \lambda_5 - \lambda_6\}, \ee

\noindent where the $\lambda_i$'s are the square roots of the
eigenvalues of $\rho\tilde{\rho}$ in descending order of magnitude.
Here $\tilde{\rho} = {\mathcal D}\rho{\mathcal D}^{-1}$, with the
operator ${\mathcal D}$ given by

\be \label{operad}
{\mathcal D} = \left( \begin{array}{cccccc}
0 & 0 & 0 & 0 & 1 & 0 \\
0 & 0 & 0 & -1 & 0 & 0 \\
0 & 0 & 1 & 0 & 0 & 0 \\
0 & -1 & 0 & 0 & 0 & 0 \\
1 & 0 & 0 & 0 & 0 & 0 \\
0 & 0 & 0 & 0 & 0 & 1
\end{array} \right) \,\, {\mathcal K},
\ee

\noindent where ${\mathcal K}$ stands for the complex conjugation
operator and (\ref{operad}) is written with respect to the total
angular momentum basis, ordered as $|2,2\ra$, $|2,1\ra$, $|2,0\ra$,
$|2,\m 1\ra$, $|2,\m 2\ra$ and $i|0,0\ra$.

In what follows we are going to consider systems comprising a given,
fixed number of identical fermions. Therefore, we are going to work
within the first quantization formalism.

\section{Entropic Entanglement Criteria for Systems of Two Identical Fermions}

\subsection{Entanglement Criteria Based on the von Neumann and the Linear
Entropies}

Let $\rho$ be a density matrix describing a quantum state
of two identical fermions and $\rho_r$ be the corresponding
single particle reduced density matrix, obtained by computing
the partial trace over one of the two particles.

If $\rho = |\psi_{sl} \rangle \langle \psi_{sl}|$, where
$|\psi_{sl}\rangle$ represents a separable pure state of the form
(\ref{slaterdet}), and

\be S_{\rm vN}[\rho] = - {\rm Tr}(\rho \ln\rho) \ee

\noindent is the von Neumann entropy of $\rho$, we have that
$S_{\rm vN}[\rho]=0$ and $S_{\rm vN}[\rho_r]=\ln 2$. That is, for
separable pure states we have $S_{\rm vN}[\rho]- S_{\rm
vN}[\rho_r] = -\ln2$. It then follows from the concavity property
of the quantum conditional entropy \cite{W78} that, for a
separable mixed state $\rho$ of the form (\ref{mixedsep}), $S_{\rm
vN}[\rho]- S_{\rm vN}[\rho_r] \ge -\ln2$. Consequently, all
separable states (pure or mixed) of a system of two identical
fermions satisfy the inequality

\be \label{voncrit} S_{\rm vN}[\rho_r] \le S_{\rm vN}[\rho] + \ln 2.
\ee

\noindent
Hence, if the quantity

\be D_{\rm vN} = S_{\rm vN}[\rho_r] - S_{\rm vN}[\rho] - \ln 2 \ee

\noindent is positive the state $\rho$ is necessarily entangled.
Indeed, in the particular case of pure states this quantity
has been used as a measure of entanglement in some applications (see,
for instance, \cite{ZZX02} and references therein).
The inequality (\ref{voncrit}) can be extended to the more general
case of systems of $N$ identical fermions. From an argument
similar to the one used to derive (\ref{voncrit}) it follows that
a separable state of $N$ fermions (that is, a state that can be
written as a statistical mixture of pure states each having the
form of single Slater determinant) satisfies the inequality

\be \label{voncriten} S_{\rm vN}[\rho_r] \le S_{\rm vN}[\rho] + \ln
N. \ee

\noindent Consequently, a state of $N$ fermions violating
inequality (\ref{voncriten}) is necessarily entangled. In the case
of pure states of $N$ fermions this entanglement criteria reduces
to one of the entanglement criteria previously discussed in \cite{PMD09}.
The special case of this criterion corresponding to pure states of
two fermions was first analyzed in \cite{GM04}. When deriving the
inequalities (\ref{voncrit}) and (\ref{voncriten}) we have used
the concavity of the quantum conditional entropy. This property is
usually discussed in connection with composite systems comprising
distinguishable subsystems. However, within the first quantization
formalism, any density matrix of two identical fermions has
mathematically also the form of a density matrix describing
distinguishable subsystems (in fact, it is just a density matrix
that happens to be expressible as a statistical mixture of
antisymmetric pure states). Consequently, any mathematical
property that is satisfied by general density matrices describing
distinguishable subsystems is also satisfied by the special subset
of density matrices that can describe a system of identical
fermions.

An entanglement criterion for states of two fermions similar to
the one already discussed  can be formulated in terms of the
linear entropy,

\be S_L[\rho] = 1 - {\rm Tr}(\rho^2). \ee

\noindent
Given a quantum state $\rho$ of two fermions, let's
consider the quantity

\be \label{cerro}
c[\rho] \, = \, \inf \sum_i p_i c[|\phi_i\rangle],
\ee

\noindent where $c[|\phi_i\rangle] = \sqrt{2\left[1-{\rm
Tr}[(\rho^{(i)}_r)^2 ]\right]}$, $\rho^{(i)}_r$ is the one
particle reduced density matrix corresponding to $|\phi_i
\rangle$, $\rho = \sum_i p_i |\phi_i \rangle \langle \phi_i |$,
and the infimum is taken over all the possible decompositions of
$\rho$ as a statistical mixture $\{p_i, |\phi_i\rangle \}$ of pure
states (note that $c[\rho]$ adopts values in the range
$[0,\sqrt{2}]$). The quantity defined in (\ref{cerro}) satisfies
the inequality \cite{MB07}

\be \label{MB} c[\rho]^2 \, \ge \, 2 \left[{\rm Tr}(\rho^2) - {\rm
Tr}\left(\rho_r^2 \right) \right]. \ee

\noindent If $\rho$ corresponds to a separable state of the two
fermions, we have that $\rho = \sum_i p_i |\psi_{sep}^{(i)}
\rangle \langle \psi_{sep}^{(i)}|$ with $c[|\psi_{sep}^{(i)}
\rangle] = 1$ for all $i$. Therefore, for a separable state we
have $c[\rho] \le 1$ and, from (\ref{MB}), $1 \ge
\left(c\left[\rho \right] \right)^2 \ge 2\left[{\rm Tr}(\rho^2) -
{\rm Tr}\left(\rho_r^2 \right) \right]$. Consequently, separable
states (pure or mixed) of a system of two identical fermions
comply with the inequality,

\be \label{lincrit}
 S_L[\rho_r] \le S_L[\rho]
+ \frac{1}{2}. \ee

\noindent In other words, states for which the quantity

\be \label{DL} D_L = S_L[\rho_r] - S_L[\rho] - \frac{1}{2} \ee

\noindent is positive are necessarily entangled. In the particular
case of pure states of two identical fermions, the  positivity of
(\ref{DL}) becomes both a  necessary and sufficient entanglement
criterion (\cite{PMD09} and references therein). 
Moreover, a quantity basically equal to (\ref{DL}) has
been proposed as an entanglement measure for pure states of two
fermions and indeed constitutes one of the most useful entanglement
measures for these states \cite{BBB07}.
\\

\subsection{Entropic Entanglement Criteria Based on the R\'enyi Entropies}

On the basis of the R\'enyi family of entropies we are going to
derive now a generalization of the separability criterion
associated with inequality (\ref{voncrit}). We are going to prove
that a (possibly mixed) quantum state $\rho$ of a system of two
identical fermions satisfying the inequality

\be\label{renycrit}
 S^{(R)}_q[\rho] + \ln 2 < S^{(R)}_q[\rho_r],
\ee

\noindent for some $q \ge 1$, is necessarily entangled. Here
$S^{(R)}_q$ stands for the R\'enyi entropy,

\be \label{riqi}
 S^{(R)}_q [\rho] = \frac{1}{1-q}\ln({\rm Tr}[\rho^q]).
\ee

\noindent The inequality (\ref{renycrit}) leads to an entropic
entanglement criterion that detects entanglement whenever the
quantity \be R_q = S^{(R)}_q[\rho_r] - S^{(R)}_q[\rho] - \ln 2 \ee

\noindent is strictly positive. In the limit $q \rightarrow 1$ the
R\'enyi measure reduces to the von Neumann entropy and we recover
the entanglement criterion given by inequality (\ref{voncrit}).
When $q \rightarrow \infty$ the R\'enyi entropy becomes

\be
 S^{(R)}_{\infty} [\rho] = -
 \ln \left( \lambda_{\rm max.} \right),
\ee

\noindent where $\lambda_{\rm max.}$ is the largest eigenvalue of
$\rho$. In this limit case, the entropic criterion says that any
state satisfying

\be   2  \, \lambda_{\rm max.}^{(\rho_r)} <  
\lambda_{\rm max.}^{(\rho)} \ee

\noindent
is entangled, where $\lambda_{\rm max.}^{(\rho)}$ and
$\lambda_{\rm max.}^{(\rho_r)}$ are, respectively, the largest
eigenvalues of $\rho $ and $ \rho_r$.

\subsection{Proof of the Entropic Criteria Based on the R\'enyi Entropies}

The following proof is based on the powerful techniques related to
the majorization concept \cite{NK01,RC03} that were introduced 
to the field of
quantum entanglement by Nielsen and Kempe in \cite{NK01}. These
authors proved that non-entangled  states of quantum systems
having distinguishable subsystems are such that the total density
matrix is always majorized by the marginal density matrix
associated with one of the subsystems. In the case of
non-entangled states of a system of identical fermions the total
density matrix $\rho $ is not necessarily majorized by the one
particle reduced density matrix $\rho_r$. However, as we are going
to prove, {\it there is still a definite majorization-related relation
between $\rho$ and $\rho_r$ that yields a family of inequalities
between the R\'enyi entropies of these two matrices, which leads
in turn to a family of entropic entanglement criteria}.

 In our proof of the entropic criterion associated 
with the inequality (\ref{renycrit}) we are
going to use the following fundamental property of quantum
statistical mixtures. If $\rho = \sum_i p_i |a_i\ra\la a_i| =
\sum_j q_j |b_j\ra\la b_j|$ are two statistical mixtures
representing the same density matrix $\rho$, then there exists a
unitary matrix $\{U_{ij}\}$ such that \cite{NK01,NC00} \be
\sqrt{p_i}|a_i\ra = \sum_j U_{ij} \sqrt{q_j}|b_j\ra. \ee \noindent

 Let us now consider a separable state of two identical fermions,

\be\label{rho} \rho = \sum_j \frac{p_j}{2}(|\psi^{(j)}_1 \ra
|\psi^{(j)}_2\ra - |\psi^{(j)}_2 \ra |\psi^{(j)}_1 \ra)(\la
\psi^{(j)}_1 | \la \psi^{(j)}_2 | - \la \psi^{(j)}_2 | \la
\psi^{(j)}_1 |) \ee

\noindent where $0 \leq p_j \leq 1$, $\sum_j p_j =1$ and
$|\psi^{(j)}_1 \ra$, $|\psi^{(j)}_2 \ra$ are normalized
single-particle
states with $\la\psi^{(j)}_1 |\psi^{(j)}_2 \ra = 0$.\\

\noindent Let us consider now a spectral representation

\be\label{rho spectral} \rho = \sum_k \lambda_k |e_k\ra\la e_k| \ee

\noindent of $\rho$. That is, the $|e_k\ra$ constitute an
orthonormal basis of eigenvectors of $\rho$ and the $\lambda_k$
are the corresponding eigenvalues. Then, (\ref{rho}) and (\ref{rho
spectral}) are two different representations of $\rho$ as a
mixture of pure states. Therefore, there is a unitary matrix $U$
with matrix elements $\{U_{kj}\}$ such that

\be\label{unitary lambda} \sqrt{\lambda_k}|e_k\ra = \sum_j U_{kj}
\sqrt{\frac{p_j}{2}}(|\psi^{(j)}_1 \ra |\psi^{(j)}_2 \ra -
|\psi^{(j)}_2 \ra |\psi^{(j)}_1 \ra). \ee

\noindent The single particle reduced density matrix corresponding
to the two fermions density matrix (\ref{rho}) is

\be\label{rhor} \rho_r = \sum_j \frac{p_j}{2}(|\psi^{(j)}_1\ra
\la\psi^{(j)}_1 | + |\psi^{(j)}_2 \ra \la\psi^{(j)}_2 |), \ee

\noindent admitting a spectral representation \be\label{rhor
spectral} \rho_r = \sum_l \alpha_l |f_l\ra\la f_l|. \ee

\noindent We now define,

\ben
q_{2j} =& q_{2j-1}& = \frac{1}{2} p_j \qquad (j=1,2,3,\hdots)\\
|\phi_{2j-1}\ra &=& |\psi^{(j)}_1 \ra \cr |\phi_{2j}\ra &=&
|\psi^{(j)}_2 \ra \qquad (j=1,2,3,\hdots). \een

\noindent Now, since (\ref{rhor}) and (\ref{rhor spectral})
correspond to two statistical mixtures yielding the same density
matrix, there must exist a unitary matrix $W$ with matrix elements
$\{W_{jl}\}$ such that,

\be\label{unitary q} \sqrt{q_i}|\phi_i\ra = \sum_l W_{il}
\sqrt{\alpha_l}|f_l\ra \qquad (i=1,2,3,\hdots). \ee

\noindent Now, eq.(\ref{unitary lambda}) can be rewritten as

\be\label{new unitary lambda} \sqrt{\lambda_k}|e_k\ra = \sum_j
U_{kj} \left(\sqrt{q_{2j-1}}|\phi_{2j-1}\ra |\phi_{2j}\ra -
\sqrt{q_{2j}}|\phi_{2j}\ra |\phi_{2j-1}\ra\right). \ee

\noindent Combining (\ref{unitary q}) and (\ref{new unitary lambda})
gives

\be \sqrt{\lambda_k}|e_k\ra
= \sum_l \left[ \sum_j U_{kj} \Bigl( W_{2j-1,l}|\phi_{2j}\ra  -
 W_{2j,l} |\phi_{2j-1}\ra \Bigr)\right] \sqrt{\alpha_l}|f_l\ra.
\ee

\noindent Therefore, since $\la e_k|e_{k'}\ra=\delta_{k k'}$ and
$\la f_l|f_{l'}\ra=\delta_{l l'}$, we have that



\be \lambda_k = \sum_l M_{kl}\alpha_l, \ee

\noindent where

\be M_{kl} \! = \! \left( \! \sum_{j'}
U_{kj'}^{*}\left\{W_{2j'-1,l}^{*}\la \phi_{2j'}| - W_{2j',l}^{*}\la
\phi_{2j'-1}|\right\}  \!\!  \right) \!\!\! \left( \!\sum_{j''}
U_{kj''}\left\{W_{2j''-1,l} | \phi_{2j''}\ra - W_{2j'',l} |
\phi_{2j''-1}\ra\right\}   \!\! \right). \ee

\noindent We now investigate the properties of the matrix $M$ with
matrix elements $\{M_{kl}\}$. First of all, we have

\be M_{kl} \geq 0, \ee

\noindent since the matrix elements of $M$ are of the form $M_{kl} =
\la \varSigma| \varSigma\ra$, with

\be |\varSigma\ra = \sum_j U_{kj} \left( W_{2j-1,l}|\phi_{2j}\ra  -
 W_{2j,l} |\phi_{2j-1}\ra \right).
\ee

\noindent We now consider the sum of the elements within a given row
or column of $M$. The sum of a row yields,

\ben \sum_k M_{kl}
&=& \sum_{j'j''} \delta_{j'j''} \left( W_{2j'-1,l}^{*}\la
\phi_{2j'}| - W_{2j',l}^{*}\la \phi_{2j'-1}| \right) \left(
W_{2j''-1,l} | \phi_{2j''}\ra - W_{2j'',l} | \phi_{2j''-1}\ra
\right) \cr &=& \sum_j\left(W_{2j-1,l}^{*} W_{2j-1,l} +
W_{2j,l}^{*} W_{2j,l}  \right) = \sum_i \left( W^{\dagger}
\right)_{li}W_{il} = 1, \een

\noindent while the sum of a column is,

\ben \sum_l M_{kl} &=& \sum_{j'j''} U_{kj'}^{*} U_{kj''} \left( \la
\phi_{2j'}|\phi_{2j''}\ra\left[\sum_l W_{2j'-1,l}^{*}
W_{2j''-1,l}\right] + \la \phi_{2j'-1}|\phi_{2j''-1}\ra\left[\sum_l
W_{2j',l}^{*} W_{2j'',l}\right] \right. \cr && \left.-\la
\phi_{2j'}|\phi_{2j''-1}\ra\left[\sum_l W_{2j'-1,l}^{*}
W_{2j'',l}\right] - \la \phi_{2j'-1}|\phi_{2j''}\ra\left[\sum_l
W_{2j',l}^{*} W_{2j''-1,l}\right] \right)\cr &=& \sum_{j'j''}
U_{kj'}^{*} U_{kj''} \left( \la \phi_{2j'}|\phi_{2j''}\ra
\delta_{j'j''} + \la \phi_{2j'-1}|\phi_{2j''-1}\ra\delta_{j'j''}
\right) \cr &=&
2 \sum_{j} \left( U^{\dagger} \right)_{jk} U_{kj} = 2. \een

\noindent When deriving the above two equations  we made use of the
unitarity of the matrices $\{U_{kj}\}$ and $\{W_{il}\}$. Summing up,
we have,

\ben \sum_k M_{kl} &=& 1 \cr \sum_l M_{kl} &=& 2. \een

\noindent We now define a new set of variables $\{
\lambda_i^{\prime} \}$ and a new matrix $M^{\prime}$ with elements
$M^{\prime}_{ij}$, respectively given by,

\ben
\lambda'_{2k-1} &=& \lambda'_{2k} \,=\, \frac{1}{2}\lambda_k \qquad (k=1,2,3,\hdots) \\
M'_{2k-1,l} &=&  M'_{2k,l} \,=\, \frac{1}{2}M_{kl} \qquad
(k=1,2,3,\hdots), \een \noindent and so we have \be\label{lambda
alpha} \lambda'_n = \sum_l M'_{nl} \alpha_l. \ee

\noindent By construction, then, we have

\ben \{\lambda_k\} &=& \{\lambda_1,\lambda_2,\lambda_3,\hdots\} \cr
\{\lambda'_n\} &=& \left\lbrace \frac{\lambda_1}{2},
\frac{\lambda_1}{2}, \frac{\lambda_2}{2}, \frac{\lambda_2}{2},
\frac{\lambda_3}{2}, \frac{\lambda_3}{2}, \hdots \right\rbrace. \een

\noindent Let us now compare the matrices $\{M_{kl}\}$ and
$\{M'_{nl}\}$. The matrix $\{M'_{nl}\}$ has twice as many rows as
$\{M_{kl}\}$, but the rows of $\{M'_{nl}\}$ can be grouped in pairs
of consecutive rows such that within each pair the rows are equal to
$1/2$ a row of $\{M_{kl}\}$. It follows that

\ben \sum_k M_{kl} = 1 \quad \Longrightarrow \quad \sum_n M'_{nl} =
1 \cr \sum_l M_{kl} = 2 \quad \Longrightarrow \quad \sum_l M'_{nl} =
1 . \een

\noindent Thus,

\be\label{doubly} \sum_n M'_{nl} = \sum_l M'_{nl} =1 \ee

\noindent and, therefore, $\{M'_{nl}\}$ is a doubly stochastic
matrix. Interpreting the $\lambda^{\prime}_n$'s and the $\alpha_l$'s
as probabilities, it follows from  (\ref{lambda alpha}) and
(\ref{doubly}) that the probability distribution $\{
\lambda^{\prime}_n \}$ is more ``mixed'' than the  probability
distribution $\{\alpha_l \}$ \cite{W78} (or, alternatively that $\{
\alpha_l \}$ majorizes  $\{ \lambda^{\prime}_n \}$ \cite{NK01}).
This, in turn, implies that for any R\'enyi entropy $S_q^{(R)}$ with
$q\ge 1$, we have

\be S_q^{(R)} [\lambda'_n] \geq S_q^{(R)} [\alpha_l]. \ee

\noindent Thus,

\ben S_q^{(R)} [\lambda'_n] &=& \frac{1}{1-q}\ln \left( 2 \sum_k
\left(\frac{\lambda_k}{2}\right)^q\right) \cr
&=& \ln 2 +
S_q^{(R)}[\lambda_k]. \een

\noindent Therefore, all separable states of the two-fermion system
comply with the inequality $S_q^{(R)}[\lambda_k] + \ln 2 \geq
S_q^{(R)}[\alpha _l]$ and since $\{\lambda_k\}$ and $\{\alpha_l\}$
are the eigenvalues of $\rho$ and $\rho_r$ respectively,

\be S_q^{(R)}[\rho] + \ln 2 \geq S_q^{(R)}[\rho_r]. \ee

\noindent The above inequality leads to an entanglement criterion
that detects entanglement when the indicator $R_q$ defined in
equation (\ref{riqi}) is strictly positive.

\section{Two-Fermion Systems with a Single Particle
Hilbert Space of Dimension Four}

Now we are going to apply our above derived entropic entanglement 
criteria to some parameterized families of two-fermion states. 
To this end, consider systems consisting of two fermions with a single
particle Hilbert space of dimension 4. In this case there is an
exact, analytical expression for the state's concurrence. It is then
possible to compare the range of parameters for which entanglement
is detected by the criteria with the exact range of parameters for
which the states under consideration are entangled. As mentioned in
Section II, in this case the two-fermions states can be mapped into
the states of two $s=\frac{2}{3}$ spins. The antisymmetric
eigenstates $|j,m\rangle$ of the total angular momentum operators
$J^2$ and $J_z$ constitute then a basis of the system's Hilbert
space. These states are $|0,0\ra$, $|2,-2\ra$, $|2,-1\ra$,
$|2,0\ra$, $|2,1\ra$, and $|2,2\ra$.

\subsection{Werner-Like States}

First we are going to consider a family of states consisting of a
mixture of the maximally entangled state $|0,0\ra$ and a totally
mixed state. These states are of form,

\be\label{werner} \rho_W = p|0,0\ra\la 0,0| + \frac{1-p}{6}\,I \ee

\noindent where $0 \le p \le 1$, and

\be I \, = \, |0,0\ra\la 0,0| + \sum_{m=-2}^{2}  |2, m\ra\la 2, m|
\ee

\noindent is the identity operator acting on the six-dimensional
Hilbert space corresponding to the two-fermion system. Evaluation of
the concurrence shows that these states are entangled when $p>0.4$.
For these states we have,

\ben D_{\rm vN}[\rho_W] &=& -\ln 2+\ln 4 - \frac{5}{6} (-1+p) \ln
\left(\frac{1-p}{6}\right)+\frac{1}{6} (1+5 p) \ln
\left(\frac{1}{6} (1+5 p)\right) \cr D_L[\rho_W] &=&
-\frac{7}{12}+\frac{5 p^2}{6}. \een

\noindent The minimum values $p_m$ of the parameter $p$ 
such that for $p > p_m$ the
entanglement indicators $D_{\rm vN}$, $D_L$, $R_2$, and
$R_{q\rightarrow\infty}$ are positive (and thus entanglement is
detected by the corresponding criteria) are given in the following
table (that is, in each case, entanglement is detected when $p$ is
larger than the listed value).


\begin{center}
  \begin{tabular}{| c | c | c | c | c |}
    \hline
              & $D_{\rm vN}>0$  & $D_L>0$
              & $R_{q\rightarrow\infty}>0$ & $R_{q=2}>0$ \\ \hline
    $p_{min}$ & $\approx 0.809$ & $\sqrt{0.7}\approx 0.837$ &  0.4
                     & $ \approx 0.632 $\\ \hline
  \end{tabular}
\end{center}

\noindent The entanglement detection efficiency of the entropic
criterion based upon R\'enyi entropy increases with $q$. Indeed,
in the limit $q\rightarrow\infty$ the R\'enyi entropic criterion
detects all the entangled states within the family of states
(\ref{werner}). The behaviour of the minimum value of $p$ for
which entanglement is detected as a function of the entropic
parameter $q$ is depicted in Figure 1.

\begin{figure}
\begin{center}
\includegraphics[scale=0.3,angle=0]{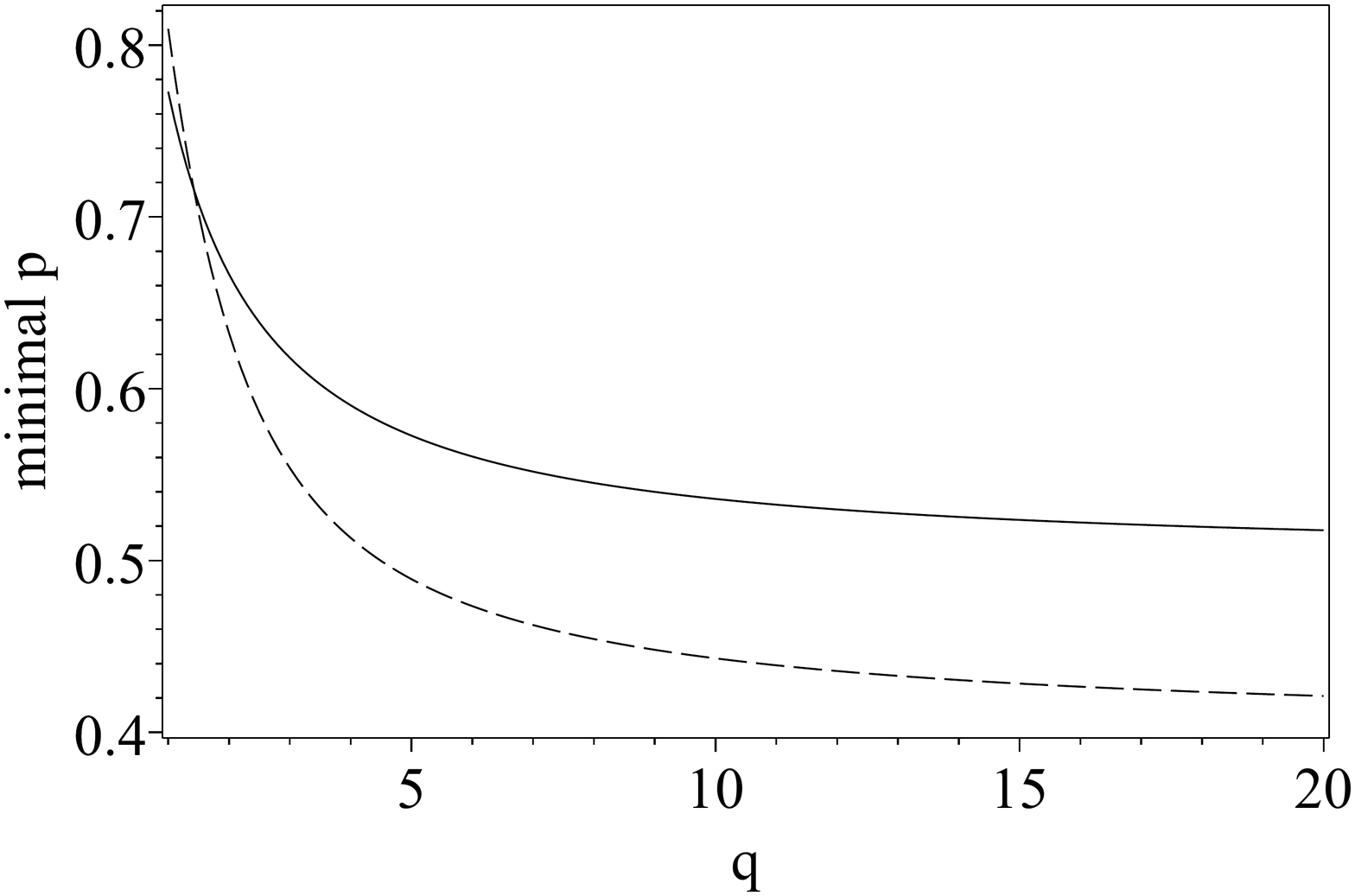}
\caption{Minimum $p$-value $p_{min}$ for which entanglement is
detected in the case of the state $\rho_W$ defined in
eq.(\ref{werner}) (dashed line) and of the state $\rho_G$ given by
eq.(\ref{gisin}) (solid line).} \label{fig1}
\end{center}
\end{figure}

\subsection{$\theta$-State}

As second illustration we consider the following pure state,

\be\label{theta} |\psi\ra =
\frac{\sin\theta}{\sqrt{2}}\left[\lli\m\frac{3}{2}\frac{3}{2}\rra
- \lli\frac{3}{2}\m\frac{3}{2}\rra\right] +
\frac{\cos\theta}{\sqrt{2}}\left[\lli\m\frac{1}{2}\frac{1}{2}\rra
- \lli\frac{1}{2}\m\frac{1}{2}\rra\right], \ee

\noindent for which

\ben D_{\rm vN}[|\psi\ra\la\psi|] &=& -\ln 2 - \cos^2\theta
\ln\left[\frac{\cos^2\theta}{2}\right]-\ln\left[\frac{\sin^2\theta}{2}\right]
\sin^2\theta \cr D_L[|\psi\ra\la\psi|] &=& \cos^2\theta
\sin^2\theta. \een

\noindent Thus, $D_{\rm vN}$, $D_L=0$ for $\theta = 0, \pi/2 ,
\pi$. So $S^{(R)}_q[\rho] + \ln 2 < S^{(R)}_q[\rho_r]$ for all
$\theta \in (0,\pi),\theta \neq \pi/2$. Therefore, all entangled
states are detected.

\subsection{Gisin-Like States}

 As a final example let us consider the parameterized family of
 mixed states given by,

\be\label{gisin} \rho_G=p|0,0\ra\la 0,0| +
\frac{1-p}{2}\,(|2,-2\ra\la2,-2| + |2,2\ra\la2,2|), \ee

\noindent with $0 \le p \le 1$. In this case we have,

\ben D_{\rm vN}[\rho_G] &=& - (-1+p) \ln (1-p) + p \ln (2 p) \cr
D_L[\rho_G] &=& \frac{1}{4} \left(-1-4 p+6 p^2\right). \een

\noindent The critical $p$ values at which the entropic criteria
based on the indicators $D_{\rm vN}$, $D_L$,
$R_{q\rightarrow\infty}$, and $R_{q=2}$ begin to detect
entanglement are listed in the Table bellow.

\begin{center}
  \begin{tabular}{| c | c | c | c | c |}
    \hline
              & $D_1>0$  & $D_2>0$      &
              $R_{q\rightarrow\infty}>0$ & $R_{q=2}>0$ \\ \hline
    $p_{min}$ & $ \approx 0.773 $ &
    $\frac{2+\sqrt{10}}{6} \approx 0.860$ & 0.5 & $\approx 0.667$ \\ \hline
  \end{tabular}
\end{center}

\noindent From the evaluation of the concurrence it follows that
the Gisin-like states are entangled for $p>0.5$. Thus, once again,
the R\'enyi based entropic criterion based on the indicator
$R_{q\rightarrow\infty}$ detects all the entangled states in the
family (\ref{gisin}).

\section{Two-Fermion Systems with a Single
Particle Hilbert Space of Dimension Six}

Two identical fermions with a 4-dimensional single particle
Hilbert space (the simplest fermionic system admitting the
phenomenon of entanglement) constitutes the only fermion system
for which an exact analytical formula for the concurrence has been
obtained. It is thus of interest to apply the entropic
entanglement criteria to systems of higher dimensionality, for
which such an expression for the concurrence is not known. Here we
are going to consider a system consisting of two identical
fermions with a single particle Hilbert space of dimension 6. The
Hilbert space of this system is 15-dimensional. Using the angular
momentum representation the two-fermion system can be mapped onto
a system of two spins with $s= \frac{5}{2}$. It is useful to
introduce the following notation,

\be \label{barritas} \left|m_1 m_2\right| =
\frac{1}{\sqrt{2}}\left[|m_1\ra |m_2\ra - |m_2\ra |m_1\ra\right].
\ee

\noindent We are going to study three particular families of mixed
states of the form

\be\label{6level} \rho_i = p|\varphi_i\ra\la \varphi_i| +
\frac{1-p}{15}\,I, \ee

\noindent where $0 \le p \le 1$ and

\be I \, = \, |0,0\ra \la 0,0| + \sum_{m=-2}^2 |2,m\ra \la 2,m| +
\sum_{m=-4}^4 |4,m\ra \la 4,m|
 \ee

 \noindent
 is the identity operator acting on the 15-dimensional Hilbert space
 describing the two-fermion system, and $|\varphi_i\ra$ is an
 entangled two-fermion pure state. We consider three
 particular instances of $|\varphi_i\ra$. In each case we provide
 the expressions for the indicators $D_{\rm vN}$ and $D_L$, and the
 minimum values $p_m$ of the parameter $p$ such that for $p>p_m$ entanglement 
 is detected by the criteria based on the positivity of the quantities
 $D_{\rm vN}$, $D_L$, $R_{q\rightarrow \infty}$ and $R_{q=2}$.


 The first illustration corresponds to
\be\label{phi1} |\varphi_1\ra =
\frac{1}{\sqrt{3}}\left[\lli\frac{5}{2}\frac{3}{2}\rli +
\lli\frac{1}{2}\m\frac{1}{2}\rli -
\lli\m\frac{3}{2}\m\frac{5}{2}\rli\right], \ee for which \ben
D_{vN}[\rho_1] &=& \ln3-\frac{14}{15} (-1+p)
\ln\left(\frac{1-p}{15}\right)+\frac{1}{15} (1+14 p)
\ln\left(\frac{1}{15} (1+14 p)\right) \cr D_L[\rho_1] &=&
\frac{1}{15} \left(-9+14 p^2\right), \een resulting in

\begin{center}
  \begin{tabular}{| c | c | c | c | c |}
    \hline
              & $D_{vN}>0$  & $D_L>0$      &
              $R_{q\rightarrow\infty}>0$ & $R_{q=2}>0$ \\ \hline
    $p_{min}$ & $\approx 0.767$ & $\frac{3}{\sqrt{14}}\approx 0.802$ &
    $\frac{2}{7}$ & $\approx 0.535$ \\ \hline
  \end{tabular}
\end{center}


\noindent The second example is given by \be\label{phi2}
|\varphi_2\ra = -\frac{2}{3}\lli\frac{5}{2}\frac{3}{2}\rli -
\frac{2}{3}\lli\frac{1}{2}\m\frac{1}{2}\rli +
\frac{1}{3}\lli\m\frac{3}{2}\m\frac{5}{2}\rli, \ee with, \ben
D_{vN}[\rho_2] &=& \frac{1}{45} \left(-45 \ln2-42 (-1+p)
\ln\left(\frac{1-p}{15}\right)+5 (-3+2 p)
\ln\left(\frac{1}{6}-\frac{p}{9}\right)-10 (3+p)
\ln\left(\frac{3+p}{18}\right) \right. \cr
              && \left. +3 (1+14 p) \ln\left(\frac{1}{15} (1+14 p)\right)\right) \cr
D_L[\rho_2] &=& -\frac{3}{5}+\frac{121 p^2}{135}, \een

\noindent and

\begin{center}
  \begin{tabular}{| c | c | c | c | c |}
    \hline
              & $D_{vN}>0$  & $D_L>0$      &
              $R_{q\rightarrow\infty}>0$ & $R_{q=2}>0$ \\ \hline
    $p_{min}$ & $\approx 0.788$ & $\frac{9}{11} $ &
    $\approx 0.324$ & $\approx 0.557$ \\ \hline
  \end{tabular}
\end{center}

\noindent
 As a third instance we tackle,

\be\label{phi3} |\varphi_3\ra =
\frac{1}{\sqrt{2}}\left[\lli\frac{5}{2}\frac{3}{2}\rli +
\lli\frac{1}{2}\m\frac{1}{2}\rli\right], \ee leading to, \ben
D_{vN}[\rho_3] &=& \frac{1}{15}\left( -p \ln7776+p \ln248832-9
(-1+p) \ln(1-p)-5 (2+p) \ln(2+p) \right. \cr
              && \left. +\ln\left(\frac{1024 (1+14 p)}{30517578125}\right)+14 p \ln(1+14 p)\right) \cr
D_L[\rho_3] &=& -\frac{3}{5}+\frac{17 p^2}{20}, \een and
\begin{center}
  \begin{tabular}{| c | c | c | c | c |}
    \hline
              & $D_{vN}>0$  & $D_L>0$      &
              $R_{q\rightarrow\infty}>0$ & $R_{q=2}>0$ \\ \hline
    $p_{min}$ & $ \approx 0.825$ & $2 \sqrt{\frac{3}{17}}\approx 0.840$
    & $\approx 0.348$ & $\approx 0.590$ \\ \hline
  \end{tabular}
\end{center}

\begin{figure}
\begin{center}
\includegraphics[scale=0.3,angle=0]{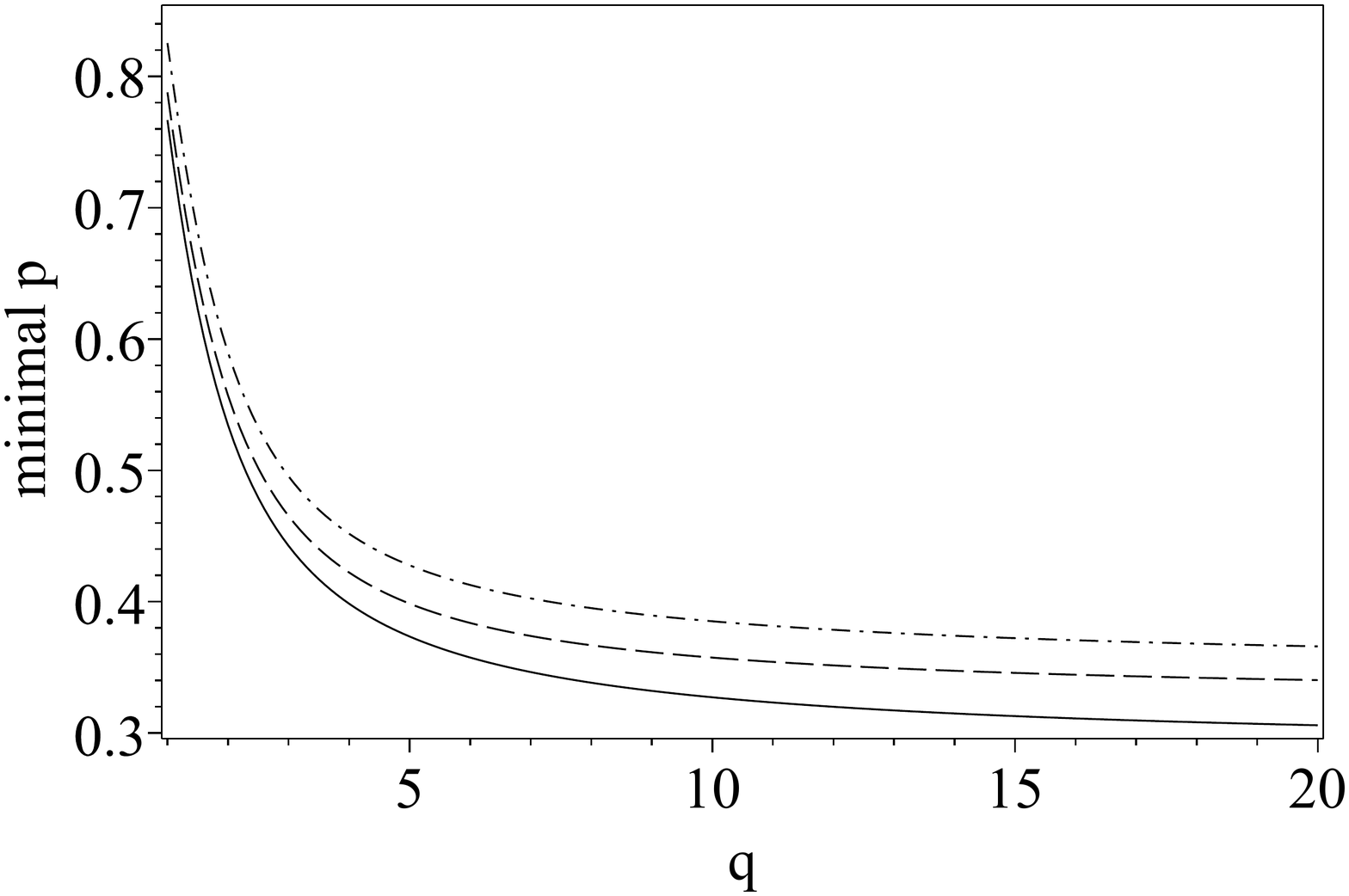}
\caption{Minimum value of  $p$, as a function of the entropic
parameter $q$, for entanglement detection in the states
(\ref{6level}) with $|\varphi_1 \ra$ (solid line), $|\varphi_2 \ra$
(dashed line) and $|\varphi_3 \ra$ (dashdotted line).} \label{fig2}
\end{center}
\end{figure}

\section{Systems of $N$ Identical Fermions}

Let us consider the general case of $N$ fermions with single
particle Hilbert space of general (even) dimension $n > N$. The
dimension of the Hilbert space associated with the $N$-fermion
system is then $d=\frac{n!}{(n-N)!N!}$. The R\'enyi based entropic
criterion for two fermions that we derived in Section III can be
extended to the case of $N$ fermions. According to the extended
criterion a state $\rho$ of N identical fermions satisfying the
inequality

\be \label{Nrenycrit} S^{(R)}_q [\rho_r] >  S^{(R)}_q [\rho] + \ln
N, \ee

\noindent for some $q\ge1$ is necessarily entangled. 
This criterion can be derived following a procedure 
similar to the one detailed in Section III for
the case of two fermions.

As an illustration of the entanglement criterion based on the
inequality (\ref{Nrenycrit}) let us consider a family of states of a
system of $N$ fermions having the form

\be\label{general werner}  p \,|\Phi\ra\la\Phi| + \frac{(1-p)}{d} \,
I_d, \ee

\noindent where $0 \le p \le 1$, $I_d$ is the identity operator
acting on the $N$-fermions Hilbert space, and the single particle
Hilbert space has dimension $n = k N $, with $k \ge 2$ integer. We
also assume that the (pure) $N$-fermion state $|\Phi\ra $ is of the
form

\be |\Phi\ra = \frac{1}{\sqrt{k}} \Bigl( |1,2,\dots, N| +
|N+1,N+2,\dots, 2N| + \ldots + |(k-1)N+1, (k-1)N+2,\dots, k N|
\Bigr),\ee

\noindent where $|i_1,i_2, \dots, i_N |$ denotes the Slater
determinant (as in equation (\ref{barritas})) constructed with $N$
different members $\{|i_1\ra, \ldots, |i_N\ra \}$ of an
orthonormal basis $\{|1\ra, \dots, |n\ra \}$ of the single
particle Hilbert space. The single particle, reduced density
matrix associated with the (pure) state $|\Phi\ra $ corresponds to
the totally mixed (single particle) state, $\frac{1}{n} I_n$,
where $I_n$ is the identity operator corresponding to the single
particle Hilbert space. On the basis of the R\'enyi entropic
criterion corresponding to $q \rightarrow \infty$ we identify as
entangled the states of the form (\ref{general werner}) satisfying
the inequality,

\be \ln n + \ln \left(p+\frac{(1-p)}{d}\right)-\ln N>0 \ee

\noindent and hence entanglement is detected for

\be p\,
> \, \frac{N \, (n-1)! - (n-N)! \, N!}{n! - (n-N)!\, N!}.
\ee


\noindent With $N$ fixed, we find that the efficiency of the
entanglement criterion grows as the dimension of the single particle
states, $n$, increases (that is, $p_{min}$ decreases with $n$).

\section{Summary}

In the present work new entropic entanglement criteria for systems
of two identical fermions have been advanced. These criteria have
the form of appropriate inequalities involving the entropy of the
density matrix associated with the total system, on the one hand,
and the entropy of the single particle reduced density matrix, on
the other one. We obtained entanglement criteria based upon the
von Neumann, the linear, and the R\'enyi entropies. The criterion
associated with the von Neumann entropy constitutes a special
instance, corresponding to the particular value $q=1$ of the
R\'enyi entropic parameter, of the more general criteria
associated with the R\'enyi family of entropies. Extensions
of these criteria to systems constituted by $N$ identical fermions
where also considered.

We applied our entanglement criteria to various illustrative examples 
of parametrized families of mixed states, and studied the dependence
of the entanglement detection efficiency on the entropic parameter 
$q$. The entanglement criterion improves as $q$ increases and is the 
most efficient in the limit $q \rightarrow \infty$.

\begin{acknowledgments}
The financial assistance of the National Research Foundation
(NRF; South African Agency) towards this research is hereby
acknowledged. Opinions expressed and conclusions arrived at,
are those of the authors and are not necessarily to be attributed
to NRF. This work was partially supported by the Projects
FQM-2445 and FQM-207 of the Junta de Andalucia (Spain),
by the MEC grant FIS2008-00781 (Spain), and by FEDER (EU).

\end{acknowledgments}

\end{document}